\documentclass[sigconf]{acmart}

\AtBeginDocument{%
  }

\begin{document}

\title{Beyond Sentiment: Examining the Role of Moral Foundations in User Engagement with News on Twitter}

 \author{Jacopo D'Ignazi}
 \affiliation{%
   \institution{Universitat Pompeu Fabra}
   \city{Barcelona}
   \country{Spain}
 }
 \affiliation{%
   \institution{ISI Foundation}
   \city{Turin}
   \country{Italy}
 }

 \author{Kyriaki Kalimeri}
 \affiliation{%
   \institution{ISI Foundation}
   \city{Turin}
   \country{Italy}
 }

 \author{Mariano G. Beiró}
 \affiliation{%
   \institution{Universidad de San Andrés}
   \city{Victoria}
   \country{Argentina}
 }
 \affiliation{%
   \institution{CONICET}
   \city{Buenos Aires}
   \country{Argentina}
 }

\renewcommand{\shortauthors}{D'Ignazi et al.}

\begin{abstract}

This study uses sentiment analysis and the Moral Foundations Theory (MFT) to characterise news content in social media and examine its association with user engagement. We employ Natural Language Processing to quantify the moral and affective linguistic markers. At the same time, we automatically define thematic macro areas of news from major U.S. news outlets and their Twitter followers (Jan 2020 - Mar 2021). By applying Non-Negative Matrix Factorisation to the obtained linguistic features we extract clusters of similar moral and affective profiles, and we identify the emotional and moral characteristics that mostly explain user engagement via regression modelling. We observe that Surprise, Trust, and Harm are crucial elements explaining user engagement and discussion length and that Twitter content from news media outlets has more explanatory power than their linked articles. We contribute with actionable findings evidencing the potential impact of employing specific moral and affective nuances in public and journalistic discourse in today's communication landscape. In particular, our results emphasise the need to balance engagement strategies with potential priming risks in our evolving media landscape.

\end{abstract}

\ccsdesc[500]{Information systems~Social networks}
\ccsdesc[500]{Human-centered computing~Empirical studies in collaborative and social computing}

\keywords{News Media, Social Networks, Moral Values, Sentiment Analysis}

\maketitle

\section{Introduction}

Journalism's impact on social issues coverage and polarization is not a one-sided equation. While journalism is essential for informing the public, holding power to account, and fostering productive conversations, it can also inadvertently contribute to polarization through biased reporting, sensationalism, and content distribution.
Social media significantly changed how news is generated and consumed, with over half of X (previously Twitter) users reporting finding their news on the platform~\cite{barbosa2022tweet}. 
Newspapers employ social media to disseminate breaking news, facilitating closer connections between journalists, readers, and the stories themselves, ultimately fostering diverse narratives within the community ~\cite{mcgregor2020twitter}.
Competing for user engagement, journals may adopt an overly sensational narrative to increase
readership~\cite{sumner2020adherence}.
Stories are presented in a way that evokes specific emotional responses (such as empathy, anger, and joy) and expresses values that align with certain morals or ideologies, influencing people's stances towards critical social issues such as mental health~\cite{Mittal_2023}, for example.

Here we assess \textit{how} a given piece of information is presented to audiences, focusing not only on the emotions employed ~\cite{doi:10.1177/0093650202250881, lecheler2013dealing}, but also on the much more understudied moral content~\cite{feinberg2015gulf,roy2021identifying}. Initial evidence shows that
moral/emotional political messages are diffused at
higher rates on social media~\cite{brady2017emotion}. 

Our broad research objective is to understand which emotional and moral elements are related to the effectiveness of a news tweet by combining sentiment analysis~\cite{mohammad2013crowdsourcing} and moral worldviews.
In particular, we are interested in:\\
\textbf{RQ1} Which are the main affective and moral elements characterising the news on Twitter?\\
\textbf{RQ2} Do these elements vary according to the news source?\\ 
\textbf{RQ3} How much do affective and moral content relate to the user engagement?\\ 
\textbf{RQ4} Are differences in affective and moral content associated with the diverse forms of user engagement?\\ 
\textbf{RQ5} Does the engagement vary according to the topic?

We focus on the prototypical emotions of joy,
sadness, anger, fear, trust, disgust, surprise~\cite{plutchik1980general}, while we operationalise morality according to the Moral Foundations Theory (MFT) ~\cite{Haidt200455}. MFT expresses the psychological basis of morality in terms of innate intuitions, defining the following five foundations: \textit{care/harm}, \textit{fairness/cheating}, \textit{loyalty/betrayal}, \textit{authority/subversion}, and \textit{purity/degradation}.

We gathered data from the Twitter platform over one year (January 2020 to March 2021), focusing on major news outlets in the United States and their readership, considering a random sample of their readership and collecting all their tweets and retweets. Such design allows for observing the interplay between the news outlets' narratives and the perception of their readership on thematic macro areas such as Politics, the Covid-19 pandemic, Climate change, and others. 
We analyze the relationship between the moral and affective content of the news outlets' tweets and the comments and discussions observed in their readership.
Moreover, we examine potential moral and affective differences in diverse macro thematic areas of news, which are extracted via label propagation of the most popular hashtag co-occurrences.
Then, we provide insights on the features most associated with user engagement based on the moral and affective elements, advancing the most recent findings on the matter~\cite{Chen2023}.

Employing Natural Language Processing (NLP) and unsupervised topic modelling techniques,  we show that the affective and moral features most associated with user engagement are Surprise, Trust, and Harm.
The length of the emergent discussion on Twitter (number of comments) is also mostly explained by the same features, demonstrating the ubiquitous nature of the particular attributes to enhance engagement.
Moreover, by examining other commonly assessed user engagement metrics, such as the number of likes, retweets, or reply counts, we notice that the effects of affective and moral characteristics remain consistent across them.
Finally, by applying non-negative matrix factorization (NMF) on the emotional and moral metrics we detected common patterns in how news were presented, and compared them across different news outlets and thematic areas.

In light of these findings, we contribute to the current state of the art by evidencing specific emotional and moral characteristics that the press incorporates in the narrative of specific macro areas. 
As we shift from an era of mass
communication to one of the echo chambers, tailored information, and microtargeting in the new media environment~\cite{cacciatore2016end}, moral guidance may have a negative impact on society, increasing the stigmatization of vulnerable populations and polarization.

Through this work, we contribute to a deeper understanding of the emotional and moral elements in the mainstream news outlets in the U.S. surrounding a wide range of topics.
Such insights can be built into systems that encourage users to challenge their understanding of
opposing views \cite{wang2022designing} aiming to reduce social polarisation.

\section{Related Literature}

In communication science, much work has been put towards the ``unsupervised'' inductive identification of clusters based upon the content of the article texts \cite{Nicholls2021}. For example,  \cite{field2018framing} offered a computational analysis of Russian news shedding light on intricate political strategies, while \cite{hubner2021did} assessed COVID-19 coverage in the early stages of the outbreak.

Appeal to emotion is a key element; \cite{burscher2016frames} employed cluster and sentiment analysis to investigate the conceptual validity of news related to the nuclear power issue, demonstrating the potential of computational methods in cluster analysis.
More recently, \cite{klein2021dynamics} analyzed European news outlets regarding migration, focusing on the sentiment evoked by images and text.

Delving into the psychological nuances of the news, the appeal to morality is fundamental to the opinion formation of the readership. 
To this extent, the MFT has been increasingly employed.
MFT is broadly adopted in computational social science, as it defines a clear taxonomy of values directly applicable to various topics.
MFT explains worldviews around a variety of social issues, such as vaccine hesitancy~\cite{kalimeri2019human,gaston2022moral} and the emergence of symbolism in resistance movements \cite{mejova2023authority} and has also proven helpful in understanding narratives and attitudes towards unemployment~\cite{bonanomi2017understanding,urbinati2020young}.

MFT was employed to characterise the general press biases~\cite{kuypers2020president}, as well as the biases of liberal and conservative journals around immigration and elections topics \cite{Mokhberian_2020}. More recently, \cite{araque2023beyond} assessed the gender inequality in the job market. In contrast, \cite{Mittal_2023} studied the mental health discourse on Twitter and News and the impact of the moral underpinnings in abating or exacerbating stigma.
In \cite{dehler2021topic}, structured topic modelling was used to uncover shifts in the emotional content regarding the discussion on renewable energies in German news media.
Beyond moral and emotional characteristics, studies on news and communication have employed unsupervised learning and community detection techniques to extract topics from news corpora~\cite{
hubner2021did}.

\section{Data Collection}

\begin{table*}[t]
\centering
\begin{tabular}{lrrrrr}
\toprule
{} &  News outlet tweets &  Avg. Conversation Size &  Total Followers & $\%$ Followers & Articles \\
\midrule
@AP             &            $5,682$ &                     $182.84$ &   $15,127,593$ &                     $0.52\%$ & $0.21\%$ \\
@CNN            &           $25,268$ &                     $188.47$ &   $53,242,242$ &                     $0.35\%$ & $78.36\%$ \\
@FoxNews        &            $1,031$ &                     $562.14$ &   $20,121,721$ &                     $0.35\%$ & $92.53\%$ \\
@Reuters        &           $33,741$ &                      $27.83$ &   $23,238,148$ &                     $0.37\%$ & $74.46\%$ \\
@TIME           &            $8,164$ &                      $43.18$ &   $18,065,949$ &                     $0.40\%$ & $53.59\%$ \\
@WSJ            &            $6,882$ &                      $51.33$ &   $18,705,760$ &                     $0.43\%$ & $0.04\%$ \\
@nytimes        &           $15,103$ &                     $151.48$ &   $46,808,154$ &                     $0.33\%$ & $90.48\%$ \\
@washingtonpost &           $12,050$ &                     $147.06$ &   $17,791,609$ &                     $0.52\%$ & $92.04\%$ \\
\bottomrule
\end{tabular}
\caption{Descriptive statistics of the collected data per news outlet. ``News outlet Tweets'' indicates the total number of tweets obtained from each news outlet, ``Avg. Conversation Size'' refers to the average size of the conversations stemming from the news outlet's tweets. ``Followers'' represents the total number of followers of the news outlet on the Twitter platform at the time of the data collection, while ``$\%$ Followers'' is the percentage of followers in our data for which we have the full Twitter activity. Finally, ``Articles'' represents the percentage of full news outlet texts that we managed to retrieve via the URL posted in their tweets. Note that AP and WSJ were under a paywall and hence the full text of the articles was not available, while FoxNews was not active on Twitter between November 8th, 2018 and March 18th 2020, and only recovered regular activity in September 2020.
}
\label{journal_statistics}
\vspace{-10pt}
\end{table*}

The data collection process was built upon the capture by ~\cite{schawe2023understanding}. 
Focusing on the U.S., the following news outlets were selected based on their popularity, substantial activity, and interaction on Twitter: The New York Times, Associated Press, Reuters, The Wall Street Journal, CNN, Time, Fox News, and The Washington Post. 
We considered a random sample of these news outlets' followers (see the Users subsection) and collected all their tweets and retweets between January 2020 and March 2021, based on the tweet id's provided by~\cite{schawe2023understanding} \footnote{The data collection was done during 2022, i.e., before restrictions in the access to the Twitter API were introduced.}.
Among the collected retweets, we individualized those originally tweeted by our news outlets of interest (News outlets subsection). We collected both the original article's text and the conversation stemming from the original tweet (see Conversations). 

\textbf{News outlets.}
We gathered tweets authored by the aforementioned news outlets and retweeted by the individuals in our user base. This resulted in a dataset of approximately 135,000 tweets referred to as ``News outlets tweets’‘. The number of collected tweets is shown in the first column of Table~\ref{journal_statistics}. Together with the tweets’ text, we collected information about the number of likes, retweets, quotes, and replies (including direct and chain replies).
We infer the coverage of the news outlets by comparing the amount of the NYT tweets to that of \cite{schawe2023understanding} which includes all the tweets authored by @NYT in 2020. We observed that, with our method, we recovered approximately $52\%$ of all the tweets of @NYT. Since the sampling distribution of the users was uniform across all news outlets, it is safe to assume a similar proportion for the other news outlets in our collection.

\textbf{Conversations.}
The conversation data includes all user interactions responding to the news outlets' tweets, encompassing direct responses to the original tweets (found in the news outlets' Twitter activity) and responses to other replies in a tree-like structure of interactions known as conversations. These conversation trees stemming from each tweet were retrieved using the Twitter API, and we report the average conversation size for each news outlet in Table \ref{journal_statistics}.

\textbf{Users.}
To better understand the behaviours and interests of the Twitter readership of each news outlet, we gathered the total Twitter activity of their followers.
Given that each news outlet has approximately 10 million followers, we selected a uniform sample of users from the entire follower group,  resulting in $\sim 10^5$ users per news outlet. We gathered these users' Twitter activity employing the Twitter API's user timeline endpoint. 
Approximately 100 million unique tweets were collected from these users, further reduced to approximately 40 million by retaining only English-language tweets.

\begin{table*}[ht!]
\renewcommand{\arraystretch}{1.1}
\centering
\begin{tabular}{lrrrrrrrr}
\toprule
{} &  @nytimes &   @AP &  @Reuters &  @WSJ &  @CNN &  @TIME &  @FoxNews &  @washingtonpost \\
\midrule
\% Covid          &     24.53 & 28.00 &     19.86 & 20.28 & 25.50 &  15.71 &      9.41 &            15.64 \\
\% Politics       &     22.42 & 25.26 &     11.89 & 15.46 & 23.80 &  16.71 &     43.74 &            25.76 \\
\% BLM            &      3.20 &  5.35 &      1.06 &  1.32 &  3.24 &   3.08 &      1.55 &             1.53 \\
\% Elections      &      5.28 &  5.25 &      1.63 &  2.76 &  5.21 &   2.57 &      6.11 &             1.52 \\
\% Climate Change  &      1.15 &  1.17 &      0.97 &  0.93 &  1.47 &   0.54 &      0.19 &             0.15 \\
\% Entertainment  &      3.75 &  2.34 &      2.68 &  3.42 &  3.76 &   4.50 &      0.68 &             2.69 \\
\% Job\&Technology &      1.85 &  1.59 &      4.09 &  6.57 &  2.93 &   1.55 &      0.48 &             1.05 \\

\bottomrule
\end{tabular}
\caption{Macro areas' coverage per news outlet. Each column represents the percentages of a news outlet's tweets that were labelled as belonging to each macro area by our trained classifier.}
\label{coverage_table}
\end{table*}

\begin{table}[ht]
    \centering
    \begin{tabular}{lcc}
        \toprule
        Macro area & Training F1 Score & Test F1 Score \\
        \midrule
        Covid & 0.45 & 0.44 \\
        Politics & 0.46 & 0.46 \\
        BLM & 0.35 & 0.34 \\
        Elections & 0.24 & 0.23 \\
        Climate Change & 0.40 & 0.39 \\
        Entertainment & 0.52 & 0.52 \\
        Job \& Technology & 0.52 & 0.52 \\
        \bottomrule
    \end{tabular}
    \caption{Average F1 score for the training and test sets (cross-validated) on each thematic macro area. }
    \label{tab:f1-scores}
    \vspace{-10pt}
\end{table}

\raggedbottom

\textbf{News Outlets' Corpus.}
For each tweet published by a news outlet in our dataset, we gathered the complete text, when available, from the respective URL links in the tweets. Table \ref{journal_statistics} reports the descriptive statistics on the obtained articles. 
Note that AP and WSJ were under a paywall, and hence the full text of the articles was not available, while FoxNews was not active on Twitter between November 8th, 2018 and March 18th, 2020 \footnote{https://variety.com/2020/tv/news/fox-news-twitter-return-1203538217/}. 

\section{Methods}

In this Section, we describe the different user engagement metrics, extracting emotional and moral characteristics from tweets and classifying news outlets' tweets into news macro areas. Finally, we describe the general regression setup used for studying the relationship between emotional/moral content and engagement.

\subsection{Engagement metrics}

Each type of engagement metric on Twitter (likes, retweets, replies, and quotes) assesses a different aspect of how users interact with the content:
\textit{likes}  can indicate the level of agreement or approval of a tweet and are often seen as a baseline engagement metric;
\textit{retweets} are a powerful metric as it reveals that other users found it valuable enough to share with their followers;
\textit{replies} are the most critical indicator of conversational engagement; \textit{quotes} indicate a more profound level of engagement, which might be aligned with or contrary to the original content.
Here, we consider all the above metrics to offer a holistic view of the engagement \cite{munoz2022measuring}.
We also estimate the average sentiment of replies under a news outlet's tweet, which can be interpreted as the general positivity/negativity induced by the news outlet in its readership~\cite{BUDER2021106663}.

\subsection{News Macro Areas}

The collected tweets cover a broad portfolio of news. For a more in-depth understanding, we aimed to classify the tweets by assigning them
a broad thematic label in a data-driven way.  
Our classification strategy builds upon the hashtags present in the tweets: as our capture is not based on a hashtag list, the tweets contain a rich variety of hashtags.  
However, as hashtags were present in only 14\% of tweets, we opted for an unsupervised labelling strategy followed by a label propagation approach. We first selected the hashtag used at least once per day on average and by at least ten authors. To identify the main thematic areas of news, we then performed a consensus clustering based on the hashtags' co-occurrences in the users' tweets. Finally, we propagated these labels by training a classifier for each thematic area.

To do so we represented each hashtag with an embedding vector, training a Word2Vec model (from Gensim Library) on a corpus where the sentences were made of the hashtags used on each tweet. 
We then applied KMeans clustering on the versors of the embedding, repeating the process 20 times for each K in the range of 1 to 100. 
The $C^K_{ij}$ element of the consensus matrix is the relative frequency with which hashtag $i$ is found in the same cluster as hashtag $j$. 
\begin{eqnarray*}
C^K_{ij}=\frac{\sum_{h=1}^{20}c^{K,h}_{ij}}{20} \, ,
\end{eqnarray*}
where $c^{K,h}_{ij} = 1$, if i is in the same cluster of j during h-st iteration of clustering for K.
As a final step, we obtained a partition of the consensus matrix for each K via KMeans, and by using the proportion of ambiguous clustering (PAC) score we obtained an optimal value of $K=30$.
After manually inspecting the resulting clusters, we merged them into ``macro areas", namely, ``Politics'', ``Job \& Technology'', ``Entertainment'', ``Climate Change'', ``Elections'' and ``Covid''.

After training the classifier on the \textit{users'} tweets that contain hashtags, we propagated the thematic macro area labels to the rest of the data (i.e., both \textit{users'} and \textit{news outlets'} tweets).
As a preprocessing step, we removed URLs, symbols, and stop words and applied tokenization and lemmatization.
To avoid overfitting, we kept only those lemmas appearing at least $500$ times in the training dataset and $1000$ times in the rest of the tweets; in this way, we kept $87.7\%$ of the corpus volume (total amount of tokens in the dataset), while using only $0.3\%$ of unique lemmas.
The dataset was thus represented as a list of BOWs and fed to an independent linear regressor for each macro thematic area.
Moreover, we applied $L1$ regularization and performed 5-fold nested cross-validation.
Since we are mainly interested in achieving satisfactory precision, we chose a threshold of $0.5$ to discretize the output of each regression model, with precision in the range of $0.7,0.75$ for each model. The resulting coverage of each thematic area is shown in Table \ref{coverage_table}.

Finally, we assessed the precision of the label-propagation step with a manual annotation of tweets by three independent annotators. We annotated a random subset of 250 tweets, containing 50 tweets from each of the macro areas Covid, Politics, Entertainment, and Job\&Technology, and 50 tweets with no macro area assigned. We obtained a Cohen's $\kappa$ of $0.62$ when assessing the agreement between the model and the annotators, while the inter-annotator agreement was $\kappa=0.67$. This latter finding warns that assigning a macro area to a tweet is not an easy task~\cite{schaefer2020annotation} even for human annotators.

\begin{figure*}[t!]%
\centering
\includegraphics[width=0.7\textwidth]{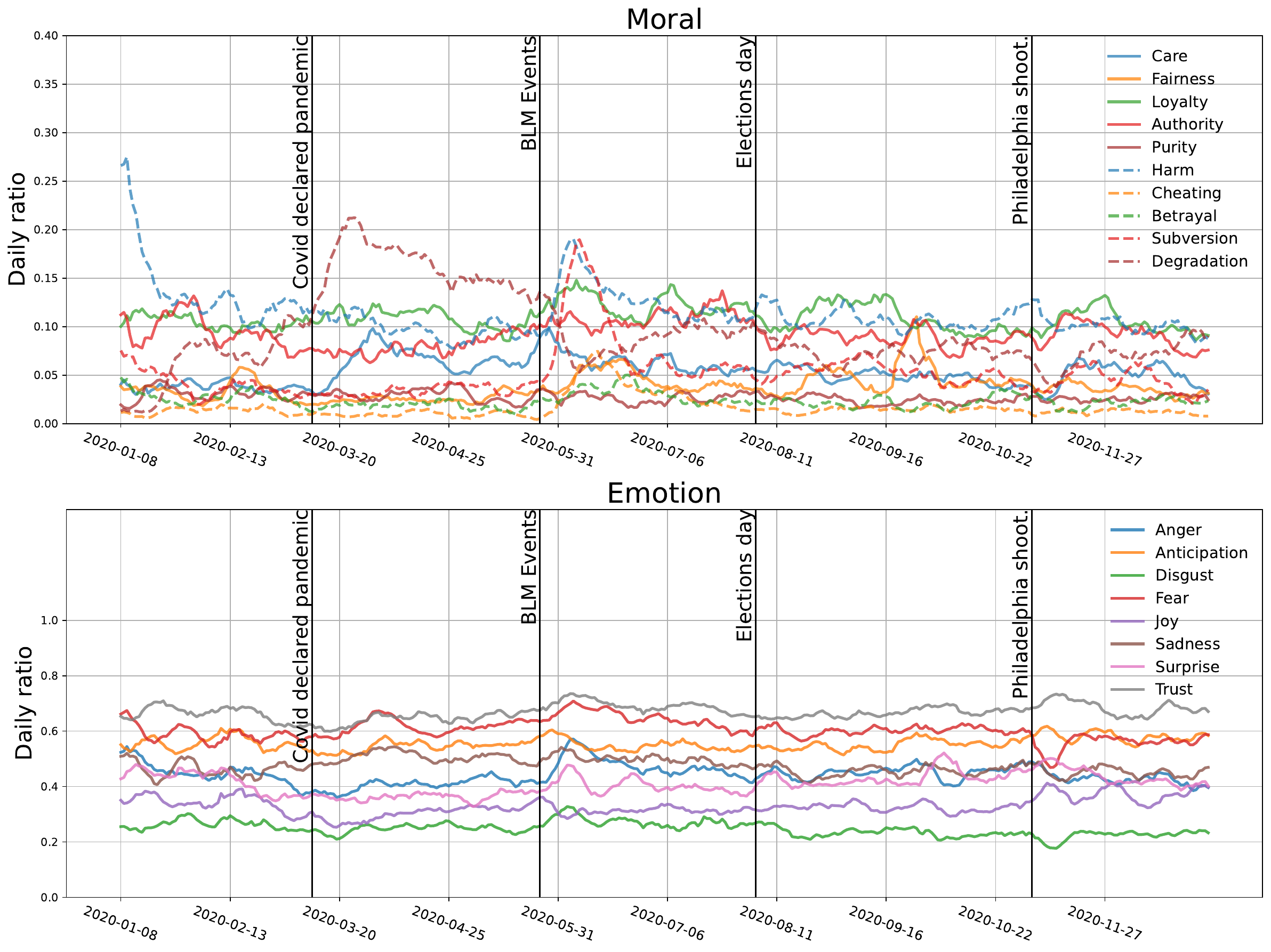}
\caption{Time evolution of the daily ratio of news outlets' tweets expressing a given moral (top) or emotion (bottom) on a moving window of $7$ days.}
\label{emotions_morals_journals}
\vspace{-10pt}
\end{figure*}

\subsection{Emotional and Moral Assessment}
\label{sec:emomoral}
In this study, we opted for a document-level lexicon-based analysis due to the approach's velocity in large-scale datasets, the fast learning process, and the direct interpretability of the results~\cite{taboada2016sentiment} \footnote{We have tested our measures both at the document-level and as a sentence-level average, observing that the results were stable across the two approaches.}.  
We employ the \textit{NCR EmoLex} \cite{mohammad2013crowdsourcing} for the emotional assessment. The EmoLex associates several lemmas with expressing eight basic emotions: anger, anticipation, disgust, fear, joy, sadness, surprise, and trust. Furthermore, it provides these lemmas with a value of $-1$, indicating a negative sentiment, or $+1$ for a positive sentiment. All the lemmas that do not appear in \textit{EmoLex} are considered neutral in the emotional and sentiment axis, thus being assigned a value of $0$. We applied an essential preprocessing step to the tweets' text, including removing non-alphanumeric characters, emojis, stopwords, and lemmatization.

For each emotion $i$ we assign a value of $1$ to the lemmas appearing in the corresponding dictionary, and $0$ otherwise. Then, the emotional presence in a tweet $t$ is computed by averaging the values assigned to each word $w$ in the set of non-stopwords of the tweet, $W_{ns}^t$, so that the tweet will be represented as a vector $E_i^t$ of values in the interval $[0,1]$:
\begin{equation}
E_i^t=\frac{\sum_{w\in W_{ns}^t}D_i^w}{|W_{ns}^t|}
\label{eq:morals}
\end{equation}
where $D_i^w\in\{0,1\}$ indicates the presence of word $w$ in dictionary associated to the emotional dimension $i$. 

To assess the moral values expressed by both the news outlets and their Twitter readership, we employ the \emph{MoralStrength} lexicon~\cite{Araque2020}. MoralStrength builds on top of the original Moral Foundations Dictionary, MFD~\cite{Graham2009} addressing its fundamental limitations such as sparsity, providing a more extensive coverage with more commonly used lemmas. 
Importantly, this lexicon provides a \textit{Moral Valence score} for several words in a dictionary associated with each moral foundation. 
Having assessed the potentials of other lexicons (e.g. \cite{Graham2009,hopp2021extended})  we opted for MoralStrength since it provides assessments for the presence of the vice and virtue loadings of each foundation. Being validated on the MFTC benchmark dataset~\cite{hoover2020moral}, the MoralStrength provides the Moral Valence score, a numeric assessment indicating the polarity and the intensity of the word in that moral foundation. In this way, the interpretation of the moral content of a 
tweet becomes straightforward. Instead, MFD~\cite{Graham2009} and eMFD~\cite{hopp2021extended}  provide one moral score for the ``care'' and one for the ``harm'' foundation, often in comparable quantities leading to ambiguous insights. 

The moral valence of a word is expressed on a Likert scale from 1 to 9, with 5 being considered neutral.
Scores lower than 5 reflect notions closer to Harm, Cheating, Betrayal, Subversion, and Degradation, while values above 5 indicate Care, Fairness, Loyalty, Authority, and Purity, respectively. 

We assess the moral content of a tweet $t$ by computing the average score of the lemmas present in the MoralStrength lexicon, $W^t_{mo}$. This gives us a vector of $m_k^t$ values in the interval $[1,9]$ for the tweet $t$ in the moral dimension $k$:
\begin{equation}
    m_k^t=\frac{\sum_{w\in W^t_{mo}}{S^w_k}}{|W^t_{mo}|}
\end{equation}
where $S^w_k \in [1,9]$ is the value assigned to the word $w$ in the MoralStrenght dictionary $k$. We assign the value of $5$ (neutral) when a moral dimension is not expressed in a tweet. Then, from this vector $m_k^t$  we extract values for the virtue and vice in each dimension:

\begin{equation}
\begin{aligned}
    Virtue_k^t &= 
    \begin{cases}
        2 \cdot \frac{m_k^t}{10}  - 1 & \text{if  } m_k^t > 5 \\
        0 & \text{otherwise}
    \end{cases} \\
    Vice_k^t &=
    \begin{cases}
        2 \cdot \left(1 - \frac{m_k^t}{10}\right) - 1 & \text{if  } m_k^t < 5 \\
        0 & \text{otherwise}
    \end{cases}
\end{aligned}
\label{eq_moralstrength}
\end{equation}

The $Virtue_k^t$ and $Vice_k^t$ moral measures are vectors with values in the range $[0,1]$. In this way, we align the output of \textit{MoralStrength} to that of \textit{EmoLex}.
 
For both the emotion and moral extraction, similar to ~\cite{,garten2018dictionaries,trager2022moral,guo2023data} we opted for a lexicon-based approach versus large language models or a pre-trained BERT model 
\cite{preniqi2024moralbert}. Lexicon-based methods can be sensitive to context and may not accurately capture the nuances of language use, however, for the scope of our analysis we chose them due to the straightforward qualitative interpretation of the results. On the other hand, pre-trained models are sensitive to the domain of training, and the generalisability of their robustness to other domains such as news is not granted ~\cite{rezapour2019enhancing}. 

\begin{table}[t!]%
\renewcommand{\arraystretch}{1.1}
\centering
\begin{tabular}{lrrrr}
\toprule
{} &  NMF1 &  NMF2 &  NMF3 &  NMF4 \\
\midrule
Anger         &  \textbf{0.705} &  0.008 &  0.049 &  0.131 \\
Anticipation  &  0.042 &  0.050 &  \textbf{0.977} &  0.281 \\
Disgust       &  \textbf{0.568} &  0.000 &  0.000 &  0.000 \\
Fear          &  0.198 &  0.000 &  0.096 &  \textbf{1.587} \\
Joy           &  0.000 &  0.000 &  \textbf{0.967} &  0.000 \\
Sadness       &  0.031 &  0.040 &  0.000 &  \textbf{1.940} \\
Surprise      &  0.000 &  \textbf{1.498} &  0.000 &  0.000 \\
Trust         &  0.107 &  0.000 &  \textbf{1.044} &  0.162 \\
Care          &  0.001 &  0.000 &  \textbf{0.117} &  0.044 \\
Fairness      &  0.038 &  0.000 &  \textbf{0.053} &  0.000 \\
Loyalty       &  0.027 &  0.000 &  \textbf{0.196} &  0.084 \\
Authority     &  0.041 &  0.000 &  \textbf{0.186} &  0.027 \\
Purity        &  0.000 &  0.002 &  \textbf{0.078} &  0.000 \\
Harm          &  \textbf{0.177} &  0.058 &  0.000 &  0.101 \\
Cheating      &  0.026 &  0.000 &  0.011 &  0.000 \\
Betrayal      &  0.033 &  0.001 &  0.013 &  0.010 \\
Subversion    &  \textbf{0.081} &  0.000 &  0.029 &  0.001 \\
Degradation   &  0.000 &  0.000 &  0.000 &  \textbf{0.567} \\
\bottomrule
\end{tabular}
\caption{Composition of each cluster obtained through NMF (columns), in terms of the weights of emotions and moral values highlighted in it.}
\label{tab:nmf_components}
\vspace{-10pt}
\end{table}

\raggedbottom

\subsection{Emotional and Moral Clusters}

We followed an unsupervised approach to uncover the affective and moral nuances adopted by news media.
This approach has been previously used in the literature to characterise the morals of news through the lens of the MFT. For example, ~\cite{10.1007/978-3-030-60975-7_16} studied the moral framing of news by measuring the similarity between the articles' content and the semantic axis of each moral dimension, as computed from the GloVe embeddings of the vices and virtues words in the MFD. Then,~\cite{reiter2021studying} applied a similar analysis for political discourse datasets, while~\cite{Mittal_2023} used k-means clustering to uncover moral clusters in news articles concerning mental health.

In this study, we opted for Non-negative Factorization (NMF, ~\cite{cichocki2006new}) to unveil news latent moral and emotional factors. NMF's non-negativity constraint offers the advantage of a straightforward interpretation of the results as positive quantities that can then be associated with the initial moral and emotional dimensions. Besides, its applicability and performance in communication research have been assessed in~\cite{greene2017exploring, Korenvcic2021topic}.

To extract moral and emotional clusters, we built a matrix $\mathbf{X} = (x_{ij})$  whose elements contain the affective and moral scores $j \in \{1, ..., m\}$ for each tweet $i \in \{1, ..., n\}$.
This results in a high-dimensional sparse matrix that we linearly decompose through NMF. 
The mixing equations can be expressed in matrix notation as:
\begin{align}
\label{eq:matrix_decomposition}
  \mathbf{X} = \mathbf{W} \mathbf{H} + \mathbf{E} \, .
\end{align}

It is worth stressing that in this representation the matrix $\mathbf{X}\in\mathbb{R}^{n\times m}$ is known, while the matrices $\mathbf{W}\in\mathbb{R}^{n\times K}$ and $\mathbf{H}\in\mathbb{R}^{K\times m}$ are unknown and determined by the NMF procedure; in the result, $\mathbf{H}$ will describe the emotional and moral composition of each of the K clusters, while $\mathbf{W}$ will represent the composition of each tweet in terms of those $K$ emotional and moral clusters.
The optimal number of clusters $K$ to be extracted was determined empirically by examining the reconstruction error for $K$ ranging from $1$ to $m$. 
We chose $K=4$ by applying the ``Elbow Rule'' to the graph of average expressed variance per component since it provided a reasonable reconstruction of the initial matrix (around $70\%$) and a good compression ratio. In Table \ref{tab:nmf_components} we show $\mathbf{H}$, i.e., the composition of each cluster in terms of the moral and emotional dimensions.

\subsection{Regression Models}

We use a linear regression framework to measure the association between moral and affective characteristics and engagement. The constructed explanatory models provided insights on the effect size of different feature groups on the engagement metrics (replies, quotes, retweets, likes). We considered each of the following three feature groups: the moral scores, the emotional presence, and finally, the latent factors emerging from the NMF decomposition in the news outlets' tweets' text. In each model, we also included features representing which news macro areas are present in that news outlet's tweet and the size of its Twitter readership to control for confounding effects, thus accounting for the intrinsic popularity bias of the news outlet and trends in Twitter conversation topics. The experimental schemes followed for five different targets: likes count, retweets count, replies count, quotes count, and overall sentiment in the induced Twitter conversation. 
In each setup, we control for two confounders: the relevance of the news outlet's tweet to each thematic area (i.e., the probability assigned by the model described in the Label Propagation subsection), and the size of its readership (i.e., amount of followers) in Twitter. The results are depicted in Figure \ref{results_generative}.

All features
were quantile-normalized, while to avoid multicollinearity, we ensured that the variance inflation factor (VIF)~\cite{james2013introduction} was not larger than $5.0$ for any feature. 
The target variables were also quantile-normalized since their distribution is heavy-tailed.

For a more in-depth analysis of the topics' influence, we repeated the process for each of the news thematic areas. We extracted subsets of the dataset according to the macro areas assigned by the regressors described in section \textit{Label Propagation}. Then, we followed the linear regression experimental design 
on each topic's subdataset, keeping the number of journal followers as a confounder. The results of each model are shown in Figure \ref{results_generative_topicwise}.   

Finally, to assess the strength of the association between the different features and engagement metrics we measured the Adjusted $R^2$ of the regression when trained with each feature group, and with all the features together.
These results are reported in Table \ref{rsq_table}.

\begin{table}[t!]%
\renewcommand{\arraystretch}{1.1}
\centering
\begin{tabular}{lcccc}
\toprule
{} &  NMF 1 &  NMF 2 &  NMF 3 &  NMF 4 \\
\midrule
Covid          & 19.62 & 16.38 & 20.85 & 43.14 \\
Politics       & 29.61 & 30.17 & 25.84 & 14.38 \\
Elections      & 19.04 & 31.02 & 41.56 &  8.39 \\
BLM            & 51.67 &  7.02 & 23.10 & 18.21 \\
Climate Change  & 31.89 & 18.56 & 31.69 & 17.87 \\
Job\&Technology & 18.58 & 14.23 & 40.30 & 26.89 \\
Entertainment  & 19.51 & 18.31 & 42.21 & 19.96 \\
Overall        & 25.23 & 19.02 & 28.68 & 27.08 \\
\bottomrule
\end{tabular}
\caption{Relative predominance of each latent factor for each thematic area.}
\label{tab:nmf_results}
\vspace{-10pt}

\end{table}

\begin{table}[t!]%
\renewcommand{\arraystretch}{1.1}
\centering
\begin{tabular}{lrrrr}
\toprule 
{} &  NMF 1 &  NMF 2 &  NMF 3 &  NMF 4 \\
\midrule
@nytimes        & 28.84 & 17.95 & 29.57 & 23.64 \\
@AP             & 31.70 & 21.16 & 22.77 & 24.37 \\
@Reuters        & 21.64 & 17.75 & 28.98 & 31.63 \\
@WSJ            & 24.93 & 17.22 & 30.35 & 27.50 \\
@CNN            & 28.16 & 15.52 & 32.74 & 23.59 \\
@TIME           & 24.10 & 17.74 & 28.98 & 29.18 \\
@FoxNews        & 23.18 & 34.04 & 22.50 & 20.27 \\
@washingtonpost & 22.53 & 30.56 & 20.29 & 26.61 \\
Overall        & 25.23 & 19.02 & 28.68 & 27.08 \\
\bottomrule
\end{tabular}
\caption{Relative predominance of each latent factor for each news outlet.}
\label{tab:nmf_results_journals}
\vspace{-10pt}
\end{table}

\section{Results \& Discussion}

\begin{figure*}[ht!]%
\centering
\includegraphics[width=0.7\textwidth]{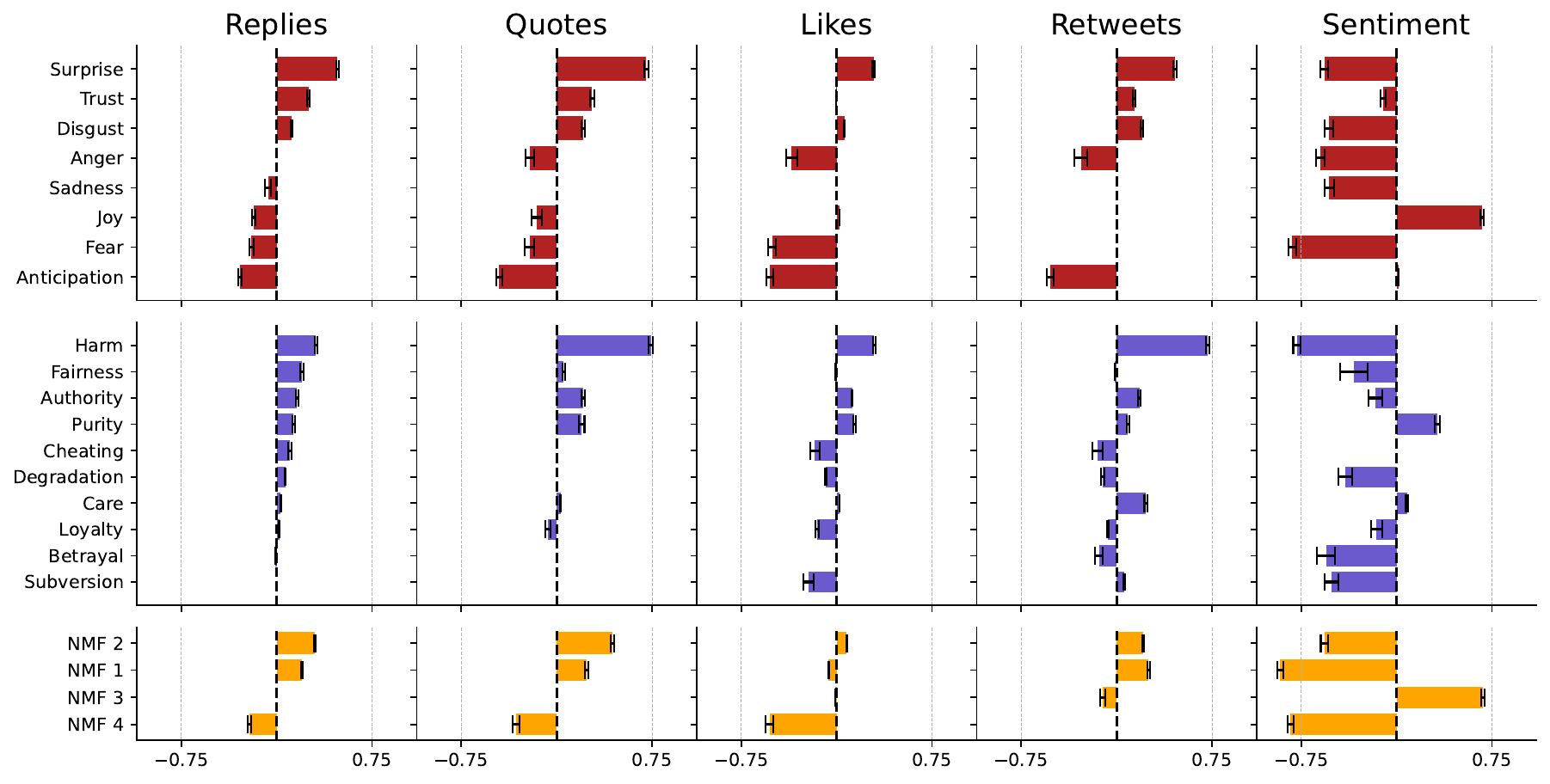}
\vspace{-10pt}

\caption{Standardized regression coefficients measuring the effect size of features extracted from the news outlets' tweets (red: only emotional features, blue: only moral features, yellow: the components of our NMF; each colour group constitutes a different experimental scheme) on different Twitter engagement metrics (replies, quotes, likes and retweets) and the sentiments found through the Twitter conversation. All the models are controlled for the number of followers of the news outlet and the tweet macro area(s) as confounders. Inside each feature group, the values are sorted according to the highest coefficient on reply count, and we only report those weights whose p-value is $\leq 0.05$.
}
\label{results_generative}
\end{figure*}

Our Twitter dataset covers the period between January 2020 and March 2021, in which the main events that monopolized public interest were the COVID-19 pandemic~\cite{paoletti2023political,lenti2023global}, the US politics~\cite{mejova2023authority} --with peaks of intense activity related to the United States presidential election--, and the death of George Floyd~\cite{mooijman2018moralization}. 
After propagating the macro area labels to the entire dataset, we noticed that news outlets covered them with roughly similar proportions (see  Table \ref{coverage_table}).
Politics and Covid are the two most present macro areas in all news outlets except FoxNews, which focuses mainly on Politics. 

The moral and affective temporal evolution in the linguistic coverage of all the above macro areas is depicted in Figure \ref{emotions_morals_journals}. We notice that Trust and Fear are the emotions with the highest prevalence, while in the expression of moral lemmas Harm, Degradation, and Subversion have the highest presence in particular after the aforementioned events.
We also notice that moral expression shows much higher variability over time concerning the time evolution of the affective lemmas.

\begin{figure*}[ht]%
\centering
\includegraphics[width=0.8\textwidth]{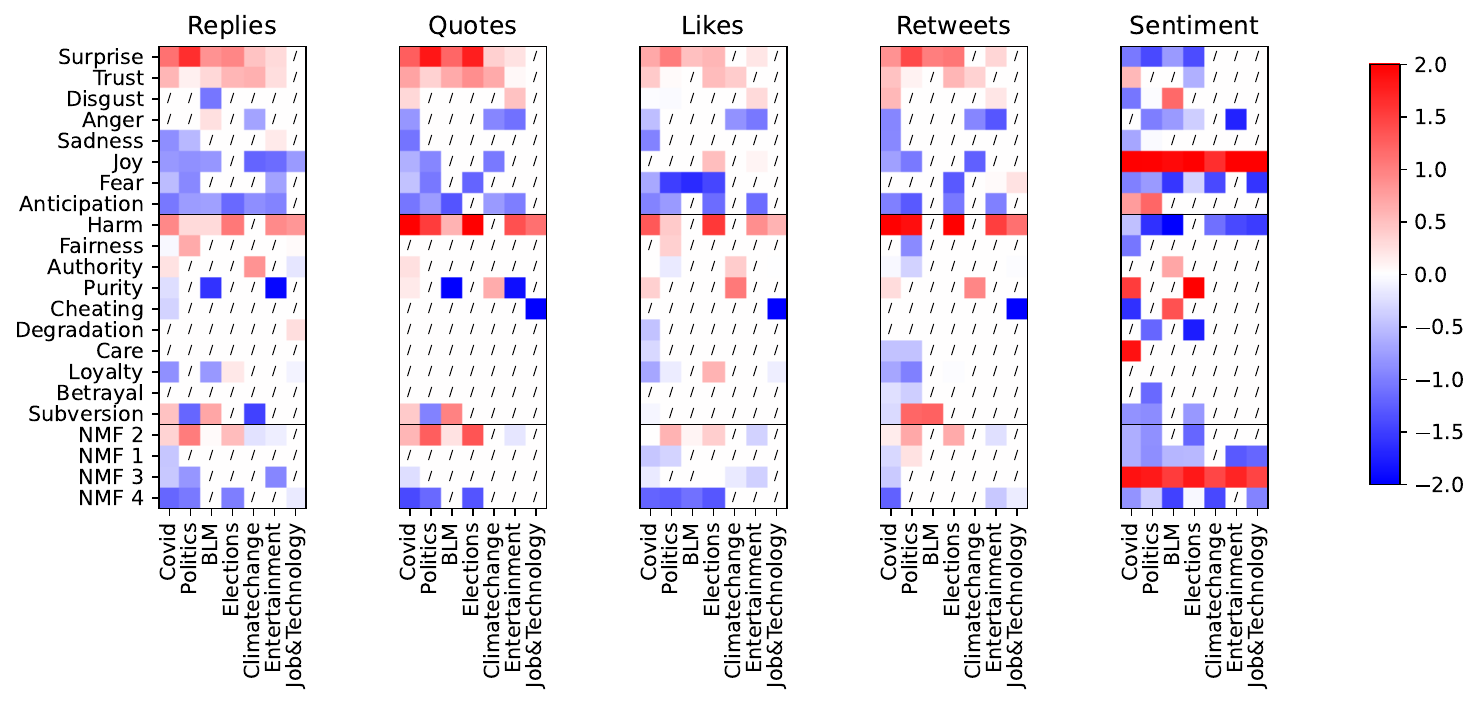}
\caption{Standardized regression coefficients measuring the effect size of emotional, moral, or NMF features on different Twitter engagement metrics,
for subsets of the news outlets' tweets corresponding to specific macro areas.
In this experiment we follow the regression setup described above on each subset of the dataset corresponding to a specific macro area. 
The horizontal bars separate different regression models. All the models are controlled for the number of followers of the news outlet, and weights whose p-value is $\geq 0.05$ are omitted and shown with a slash.
}
\label{results_generative_topicwise}
\end{figure*}

\begin{table}[ht]%
\centering
\renewcommand{\arraystretch}{1.1}
\begin{tabular}{lrrrrr}
\toprule
{} &  Rep &  Q &  L &  Ret &  S \\
\toprule
All features                       &   \textbf{.300} &  \textbf{.134} & \textbf{.212} &    \textbf{.157} &     \textbf{.106} \\
\midrule
$\#$ Followers                 &   .083 &  .043 & .086 &    .046 &     .001 \\
Macro area (T)           &   .170 &  .056 & .076 &    .070 &     .053 \\
\midrule
{Clusters (T)}          &   .051 &  .025 & .033 &    .036 &     .046 \\
\midrule
Emotions  (T)        &   .062 &  .029 & .030 &    .030 &     .060 \\
Emotions (A)      &   .056 &  .017 & .022 &    .015 &     .054 \\
\midrule
Morals (T)         &    .030 &   .016 &  .026 &     .032 &     .013 \\
Morals (A)       &    .030 &   .006 &  .016 &     .005 &     .012 \\
\bottomrule
\end{tabular}
\caption{Adjusted $R^2$ values for different linear regression models (one per row) in which our $4$ engagement metrics for a news outlet's tweet (likes (L), retweet count (Ret), reply count (Rep), and quote count(Q)) are fitted on the emotions and morals of either the content of the tweet (T) or the article (A) linked in the tweet -when available-, by the number of followers of the corresponding news outlet, and by the macro area(s) present in that tweet. The first row shows the adjusted $R^2$ for a linear regression on all the feature groups together. 
}
\label{rsq_table}
\vspace{-10pt}
\end{table}

\textbf{RQ1: Which are the main affective and moral elements characterising the news on Twitter?} To address our first research question, we employed NMF to extract latent components of the moral and affective features in the text. We report the composition of the latent factors, i.e., the latent factors, in Table \ref{tab:nmf_components}. 
For a more in-depth understanding of the different components of the same news macro area, here we provide some indicative examples. For instance, with respect to news tweets referring to the COVID-19 pandemic, we present an indicative tweet per each latent factor.

The first factor (NMF1) consists mainly of Anger, Disgust, Fear, and Harm; this factor could express social unrest, or opposition to any political or public health argument (e.g., ``U.S. envoy blames China for endangering world with coronavirus'',). The second factor (NMF2) is mainly expressing Surprise (e.g., ``BREAKING: China reports 121 more deaths, 5,090 new cases in virus outbreak, raising the death toll to nearly 1,400.''). 
The third factor (NMF3) combines Trust, Joy, Anticipation, Loyalty, and Authority (e.g.,  ``Opinion: President Trump gathers friends to reassure him he is doing a good job with the coronavirus''). Finally, the fourth factor (NMF4) consists mainly of Sadness, Fear and Degradation (e.g., ``Virus-killing robot zaps airport viruses as pandemic travel picks up'').

We estimated the relative predominance of each cluster to the macro areas (see Table~\ref{tab:nmf_results}), and we noticed that news outlet tweets employ diverse affective and moral characteristics in their reporting according to the topic.
Striking is the case of BLM, where the social unrest latent factor (NMF1) prevails (notions of anger, disgust, fear, harm, and subversion), while its second most relevant latent factor is related to trust and joy (NMF3). In social unrest movements, even though anger and disgust are the most common emotions to encourage participation in the initial period, the literature confirms that joy, trust, and anticipation can sustain people's involvement~\cite{field2022analysis, huang2023comparing}. The prevalence of harm and subversion around BLM discussions was also observed in~\cite{priniski2021mapping}. This is also observed in the time evolution present in Figure~\ref{emotions_morals_journals}, which shows a peak in these moral values just one week after the initiation of the manifestations.
The COVID-19 pandemic, instead, is mainly presented uniquely, with a particular emphasis on fear, sadness and degradation (NMF4). Again, Figure~\ref{emotions_morals_journals} shows a sharp increase in the degradation of moral value when COVID-19 is declared a pandemic; this is in line with findings in~\cite{aslam2020sentiments}, who analyzed $140$k news headlines during the first months of the pandemics. 
Finally, the most commonly adopted approach incorporates anticipation, joy, loyalty and trust (NMF4) and is mainly employed when presenting entertainment and technological news.

\textbf{RQ2: Do the moral and affective characteristics vary according to the news source?} To assess whether the predominant moral and affective characteristics vary according to the news source, we estimated the factors' prevalence per source. As can be seen in Table \ref{tab:nmf_results_journals}, we notice a general agreement 
in moral and affective style across news outlets, with mild differences due to the specific focus on the topics of interest per source.
Foxnews, for instance, employs the NMF2 latent factor which emphasizes Surpise, while Reuters and TIME prefer the NMF4 which is centred around Fear and Sadness. The
CNN, the NYT, and the WSJ, commonly employ the NMF3 which expresses notions of Trust, Joy and Anticipation, as well as the moral elements of Loyalty, Authority and Fairness. 

\textbf{RQ3: How much do affective and moral content relate to user engagement?} 
We model the user engagement, considering a series of engagement metrics (i.e., \textit{like count, retweet count, reply count, and quote count)} via linear regression.
Table \ref{rsq_table} reports the adjusted $R^2$ for each model; notice that each row corresponds to a separate regression model.
As expected when all features are considered the adjusted $R^2$ is the highest for all the target variables. 
The two confounding factors we consider, namely, the size of readership and the news macro area are the factors that explain most of the user engagement. 
Following, the proposed latent factors can explain better the Likes and Retweets, while emotions are better explainers of the number of replies and quotes. This shows that the latent factors are capturing better notions of approval, while emotions are more indicative of more profound forms of engagement~\cite{lazarus1991emotion}.
We observe that the models trained on features extracted solely from the tweet are much more likely to engage users, across all metrics than the models considering features of the linked article. This finding suggests that the brief manner news is presented on Twitter is more engaging.

\textbf{RQ4: Are differences in affective and moral content associated with the diverse forms of user engagement?} To understand which affective and moral content resonates more with our user engagement metrics, in Figure \ref{results_generative}, we depict the standardized regression coefficients obtained, controlling for the macro area of the tweet and the news outlets' number of followers. 
We observe that the NMF1, consisting of Anger, Disgust, and Harm, is consistently related to a positive increase in all the engagement metrics. The NMF2, mostly expressing excitement, follows the same pattern but with a slightly negative relationship to Likes. NMF4, consisting of Fear, Sadness and Degradation, is related to shorter engagement in terms of replies, quotes, and likes.
Zooming in the separate elements, most forms of engagement are mainly explained by the presence of Surprise, Trust, and Harm. An association between trust and surprise and user engagement had also been observed by~\cite{rathnayake2021repurposing} in the context of Covid-19. 
On the contrary, Fear,  and Anticipation are associated with less engagement. By dividing the targets between those representing endorsement (likes and retweets) and those representing an interaction (replies and quotes), we notice that Joy is associated with less interaction.

Finally, to complement this analysis, we explored the association between the moral and emotional content in news outlets' tweets, the engagement arising from them, and the overall sentiment of that Twitter conversation. 
As shown in Figure \ref{results_generative_topicwise}, users' sentiment is also associated with the news outlet tweets' moral and emotional content: tweets centred around Anger and Disgust (NMF1) are related to more negative overall sentiment, followed by the tweets containing with Fear, Degradation, and Sadness (NMF4). On the other hand, tweets expressing Joy (NMF3) are the only ones associated with positive sentiment in the conversation following the news tweet.
Interestingly, we find that moral triggering is generally associated with negative sentiment in the tweet's discussion, in line with the results of~\cite{Robertson2023}.

\textbf{RQ5: Does the engagement vary according to the topic?}  
Focusing on each specific macro area, we observe that the main morals and emotions associated with high user engagement are consistent across the various metrics, although interesting patterns emerge (see Figure \ref{results_generative_topicwise}).  
For instance,
Anger is associated with fewer quotes, likes, and retweets for Covid and climate change, however,  has more replies on BLM news tweets. 
Disgust 
is associated to fewer number of replies for the BLM. Disgust for the macro area of Covid is related to more retweets. 
Climate Change tweets are more liked when the message contains notions of purity. 
Interestingly, Fairness is related to more replies and likes but fewer retweets for political topics.
Another interesting observation is about Subversion which it relates to fewer replies and quotes but more retweets for the macro area of Politics.

\section{Conclusions}

This study provides valuable insights into the way moral and affective news content on Twitter during significant events is associated with user engagement.
News published on Twitter from January 2020 to March 2021 was dominated by significant events like the COVID-19 pandemic, U.S. politics, the United States presidential election, and the death of George Floyd (BLM movement). 
We show evidence that mainstream news outlets exhibited relatively homogeneous coverage of these events, with politics and COVID-19 being the most prevalent topics. 

Addressing our research questions, we show that the way news is expressed on Twitter significantly relates to user engagement. Emotions were more indicative of deeper forms of engagement like replies and quotes, while the approval metrics such as likes and retweets were more associated with the latent factors which combined moral and affective content. Notably, tweets that were concise and focused on the content itself tended to engage users more effectively than those linking to external articles.
While previous works employing computational approaches to news expression focused either on emotions or morals, here we combine both to provide a better understanding of social media users’ behaviour concerning news, inspired by ~\cite{brady2017emotion}. 

Different affective and moral combinations were associated with varying levels of user engagement and sentiment. 
The first latent factor reflects social unrest (BLM), the second one is composed of surprise (Elections), the third is linked to more optimistic news aspects (Entertainment), and the fourth is associated with the pandemic (Covid).
The variation in moral and affective styles among news is related to the specific topic of focus.
Interestingly, the sentiment of Twitter conversations stemming out of tweets from news outlets, was in line with the moral and emotional content of the initial tweets, with negative sentiments often associated with moral content.

Our analysis of explanatory models for user engagement revealed that Surprise, Trust, and Harm played crucial roles in explaining various forms of user engagement. At the same time, Fear and Anticipation were associated with lower engagement.
When examining specific thematic macro areas, we found that the main morals and emotions associated with engagement remained consistent, albeit with nuanced patterns. For instance, Subversion was associated with more replies to Black Lives Matter (BLM) tweets, while Climate Change tweets received more interactions when centred around purity.
Finally, sentiment analysis of Twitter conversations showed that tweets containing  Anger and Fear had a more negative overall sentiment, while tweets expressing Joy had a higher overall sentiment. Moral notions expressing the ``Harm'' foundation, were associated with negative sentiment in tweet discussions.

Controversial social issues, like public health~\cite{lenti2023global} and political and socio-economic biases~\cite{Mokhberian_2020,elejalde2018nature} play a fundamental role in promoting prosocial behaviours and identifying early symbolism partisanship~\cite{mejova2023authority}. Given the politicisation of the medium, we believe that the highlighted insights contribute to a deeper understanding of the complex interplay between news coverage, and the moral and emotional triggering of user engagement in media.

\section{Acknowledgments}

JI is part of Maria de Maeztu Units of Excellence Programme CEX2021-001195-M, funded by MICIU/AEI /10.13039/501100011033. KK and JI acknowledge support from the Lagrange Project of the ISI Foundation, funded by CRT Foundation. MGB acknowledges  support from Universidad de San Andrés.

\bibliographystyle{plain}
\bibliography{bibliography,aaai22}

\end{document}